\documentclass[prd,twocolumn,tightenlines,nofootinbib,superscriptaddress]{revtex4-1}%superscriptaddress, nofootinbib
\usepackage{bm,dcolumn,amsmath,graphicx,amsfonts,amssymb}
\usepackage{feynmp}
\usepackage{hyperref} 
\usepackage{xcolor}
\usepackage{physics}
\definecolor{cite}{rgb}{0.,0.,0.85}   % more subtle than the default blue
\hypersetup{colorlinks,linkcolor={cite},citecolor={cite},urlcolor={cite}}

\usepackage[export]{adjustbox}

\renewcommand{\v}[1]{\ensuremath{\boldsymbol{#1}}}		%bold-math for vectors
\newcommand{\be}{\begin{equation}}
\newcommand{\ee}{\end{equation}}

\def\en{\ensuremath{\varepsilon}}

\usepackage[normalem]{ulem}
\definecolor{newc}{rgb}{0.,0.6,0.4}

\begin{document} 

\title{Accurate electron-recoil ionization factors for dark matter direct detection\\ in xenon, krypton and argon}

\author{A.\ R.\ Caddell}\email[]{a.caddell@uq.edu.au}
		\affiliation{School of Mathematics and Physics, The University of Queensland, QLD 4072, Australia}

\author{V.\ V.\ Flambaum}
		\affiliation{School of Physics, University of New South Wales, Sydney NSW 2052, Australia}

\author{B.\ M.\ Roberts}\email[]{b.roberts@uq.edu.au}
		\affiliation{School of Mathematics and Physics, The University of Queensland, QLD 4072, Australia}
\date{ \today }  

%-------------------------------------------------------------------------
\begin{abstract}

    While most scintillation-based dark matter experiments search for Weakly Interacting Massive Particles (WIMPs), a sub-GeV WIMP-like particle may also be detectable in these experiments. While dark matter of this type and scale would not leave appreciable nuclear recoil signals, it may instead induce ionization of atomic electrons. Accurate modelling of the atomic wavefunctions is key to investigating this possibility, with incorrect treatment leading to a large suppression in the atomic excitation factors. We have calculated these atomic factors for argon, krypton and xenon and present the tabulated results for use with a range of dark matter models. This is made possible by the separability of the atomic and dark matter form factor, allowing the atomic factors to be calculated for general couplings; we include tables for vector, scalar, pseudovector, and pseudoscalar electron couplings. Additionally, we calculate electron impact total ionization cross sections for xenon using the tabulated results as a test of accuracy. Lastly, we provide an example calculation of the event rate for dark matter scattering on electrons in XENON1T and show that these calculations depend heavily on how the low-energy response of the detector is modelled.

\end{abstract} 

\maketitle

%======================================================
\section{Introduction} \label{sec:introduction}
%======================================================

    As the astrophysical evidence for the existence of dark matter (DM) has strengthened,
    research into its identity has started to seep into many fields of physics. However, even after years of dedicated experiments, the nature of DM remains a mystery, with no confirmed detection to date \cite{Schumann2019DirectStatus}. 
    The widely held explanation is that DM is an undiscovered particle that interacts primarily via gravity, with a very weak coupling to ordinary matter that could be exploited as a detection route \cite{Bertone2005ParticleConstraints}. Of the particle candidates, the Weakly Interacting Massive Particle (WIMP) is the most sought, with numerous experiments designed to search for WIMPs with GeV mass-scales and above (see, e.g., Refs.~\cite{Aprile2022SearchXENONnT,Aalbers2022FirstExperiment,Zhang2022SearchPandaX-4T,Agnes2018Low-MassExperiment}).

    One lesser researched option is a WIMP-like particle with sub-GeV scale mass. Many of the recent DM experiments are scintillation-based, and rely on nuclear recoil to claim a detection.
    As these lighter particles are at masses comparable to or less than nucleons, the nuclear recoil rate is negligibly small. 
    However, interactions with the atomic electrons in the scintillator may result in observable ionization signals, see, e.g., Refs.~\cite{Kopp2009DAMA/LIBRAMatter,Essig2012DirectMatter,Roberts2016DarkSignal,Roberts2019Electron-interactingDetectors,Emken2022ElectronMatter,Bloch2021ExploringExperiments,Hamaide2023FuelingCounters,Clarke2017MirrorXENON1T}.
    Low-mass WIMPs may also cause observable signals from nuclear scattering via the Migdal effect, see, e.g., Refs.~\cite{Cox2023PreciseEffect,Essig2019,Flambaum2020NewEffects,Bell2020MigdalScattering,Baxter2019,Dolan2018}.
    
    Experiments utilising dual-phase time projection chambers (TPCs) are of particular interest. In these detectors, the bulk of the scintillating material is in a liquid phase with an applied electric field (sometimes called the `drift' field), while the remaining material is in its gaseous phase above it, with a stronger electric field \cite{Aprile2017TheExperiment}. Due to this set up, results from these types of detectors can come in the form of S1 and S2 signals.
    
    The prompt scintillation signal, S1, occurs in the liquid phase when a collision between an incoming particle and an atom of the scintillator causes a release of photons. Due to the electric field in this section, if the collision instead results in atomic electrons being ionized, those electrons will drift upward, toward the gaseous phase 
    (for more details, see, e.g., Refs.~\cite{Aprile2017TheExperiment,Agnes2018Low-MassExperiment,Zhang2022SearchPandaX-4T,Liu2017CurrentDetectionexperiments}).
    For the case of sub-GeV DM interacting with a scintillator, most research focuses on looking at the S2 signal (see, e.g., Refs.~\cite{Aprile2019LightXENON1T,Cheng2021SearchExperiment}). However, DM-electron interactions may have a higher chance of creating a detectable S1 signal than previous research suggests due to enhancements in the event rate \cite{Roberts2019Electron-interactingDetectors}.

    To explore the possibility of atomic ionization we need to calculate atomic ionization factors, the details of which are discussed in Sec.~\ref{sec:calculations}. However, the calculations present many difficulties.
    Depending on the details of the experiment, accurate atomic ionization factors are often required across many orders of magnitude of energy deposition ($\sim$\,eV to keV) and momentum transfer ($\sim$\,keV to MeV). Futhermore, as inaccurate description of the atomic wavefunctions can lead to errors of up to several orders of magnitude in the calculations, this prevents many common and convenient approximations from being used, as previously discussed in Ref.~\cite{Roberts2019Electron-interactingDetectors}.

    At high values of energy and momentum transfer, relativistic effects become crucial to the calculations~\cite{Roberts2016IonizationMatter,Roberts2016DarkSignal}. These effects can even dominate the calculations, as the parts of the the electron wavefunctions that are closest to the nucleus contribute the most to scattering, and so we need fully relativistic wavefunctions to accurately model the small-distance behaviour.

    At moderate momentum transfer values, the small-distance scaling of the atomic wavefunctions is again very important. This can lead to drastic errors in the calculations when the wavefunctions are approximated as hydrogen-like and scaled by the relevant factor (sometimes referred to as ``effective-Z'' methods).

    At low momentum transfer, but also arising at all scales to some degree, the form of the continuum wavefunctions is crucial to the calculations. Approximating these continuum wavefunctions as plane waves misses the significant Sommerfeld-like enhancement~Refs.~\cite{Essig2012DirectMatter,Roberts2019Electron-interactingDetectors}. The attractive potential of the nucleus means that plane waves do not have appropriate small-distance scaling in this area. For the continuum states to be unbound energy eigenstates, they must be found in a self-consistent atomic potential to ensure correct orthogonality to the bound electrons~\cite{Tan2021,Flambaum2020NewEffects}.

    The need for accuracy also extends to the continuum state energy, as solving the Dirac equation in this region can be numerically unstable. Finally, at moderately low values of both energy deposition and momentum transfer, the atomic ionization factor can depend significantly on the atomic potential itself.

    Our method applies the relativistic Hartree-Fock approximation, and accounts for the most important many-body effects. This approach addresses all of the above issues, allowing accurate calculation at the energy deposition and momentum transfer values relevant to DM-electron scattering. Additionally, we can test our method by calculation electron-impact ionization cross sections, which is a similar process to DM-impact ionization for low-mass WIMP-like particles. High accuracy experimental rates have been measured for xenon in the relevant impact energy regime, allowing a stringent test of the accuracy of our method. In the important of $\sim$\,keV incident energies, our calculations agree with experiment substantially better than dedicated calculations that focused solely on electron impact rates.

    We present tables of atomic excitation factors for argon, krypton, and xenon as supplementary materials \cite{SeeFactors} and on GitHub \cite{Roberts2023AtomicIonisation}. These tables can be used in conjunction with DM form factors to calculate cross section and event rates, without the underestimates that arise from inaccurate atomic physics. An example of this process is provided as code, also on Github alongside the tables \cite{Roberts2023AtomicIonisation}.
    Finally, we note that our code and technique, while developed for scattering, also applies for absorption (see, e.g., Refs.~\cite{Tan2021,Derevianko2010Axio-electricEffect,Dror2021,*Dror2020c}), and may be beneficial in that case also.

%======================================================
\section{Theory} \label{sec:theory}
%======================================================

    We consider inelastic scattering between a non-relativistic DM particle and an atom. We may model the DM-electron interaction with an effective Yukawa coupling,
    \begin{equation}
        V = \hbar c \alpha_{\chi} \frac{e^{-\mu r}}{r} \, ,
    \label{eqn:yukawaV}
    \end{equation}
    where $\alpha_{\chi}$ is the DM-electron coupling strength, and $\mu$ is the inverse interaction length scale. This form of potential will result from the non-relativistic limit of a vector or scalar coupling between electrons and DM particles, in which case $\mu$ may be recognised as $\mu = m_v c / \hbar$, where $m_v$ is the mass of the mediator particle. This potential will reduce to a Coulomb-like potential if the mediator is massless, or to a contact interaction if it is heavy.

    The differential cross section for an atomic electron in initial state $n \kappa m$ to be ionized into final state $\varepsilon \kappa' m'$, can be written as
    \begin{equation}
        \frac{\mathrm{d} \sigma_{n \kappa}}{\mathrm{d} E} =
         8 \pi \alpha_{\chi}^2 \left( \frac{c}{v} \right)^2 \int_{q_-}^{q_+}
         \frac{ q \mathrm{d} q }{ \left( q^2 + \mu^2 \right)^2 } 
         \frac{ K_{n \kappa} \left( E,q \right) }{ E_H } \, .
    \label{eqn:dsnkdE}
    \end{equation}
    where $E$ is the energy deposition, $v$ is a fixed DM velocity, $q$ is the momentum transfer, $E_H \equiv m_e c^2 \alpha^2 \approx 27.2$ eV is the Hartree energy unit, which we introduce to to make $K$ dimensionless, $K$ is the dimensionless atomic excitation factor~\cite{Roberts2019Electron-interactingDetectors}, $n$ is the principal quantum number of the bound electron, and $\varepsilon$ is the energy of the ionized electron, which we can write as $\varepsilon = E - I_{n \kappa}$, with $I_{n\kappa}$ denoting the ionization energy of state $n \kappa$. The integration limits, $q_{\pm}$, are the allowed range of momentum transfer\footnote{Technically, it is $\hbar q$ that is the {\em momentum} transfer, $q$ has units of inverse length. We refer to $q$ as the momentum transfer for brevity. We keep factors of $c$ and $\hbar$ in equations to aid in comparison between works.}
    \begin{equation}
        \hbar q_{\pm} = m_{\chi} v \pm \sqrt{m_{\chi} v - 2 m_{\chi} E} \, .
    \end{equation}
    Lastly, $\kappa$ is the Dirac quantum number, defined in terms of the quantum numbers for orbital angular momentum, $l$, and total angular momentum, $j$:\ $\kappa = (l - j) (2j + 1)$.

    The atomic excitation factor is proportional to the chance of the transition from state $n \kappa m$ to $\varepsilon \kappa' m'$ occurring due to the interaction with a DM particle, and is defined in Ref.\ \cite{Roberts2019Electron-interactingDetectors} for a vector electron coupling as
    \begin{equation}
        K_{n\kappa} \equiv
         E_H \sum_{m} \sum_{\kappa' m'} \abs{\mel{ \varepsilon \kappa'm' }{e^{i \boldsymbol{q} \cdot \boldsymbol{r}}}{n\kappa m}}^2 \, ,
    \label{eqn:Knk_mel}
    \end{equation}
    where $m$ is the magnetic quantum number.
    We stress that the final states are {\em energy} eigenstates, not momentum eigenstates. It is therefore natural to normalise on the energy scale, such that
    \begin{equation}\label{eq:e-norm}
        \int\limits_{\varepsilon-\delta}^{\varepsilon+\delta} \braket{\varepsilon'\kappa'm'}{\varepsilon\kappa m}\,{\rm d}\varepsilon' = \delta_{\kappa,\kappa'}\delta_{m,m'}.
    \end{equation}
    The continuum wavefunctions then have dimension: $\left[ \phi_{\varepsilon} \right] = L^{-3/2} E^{-1/2}$. The factor $E_H$ is introduced to make $K$ dimensionless. Care should be taken comparing ionization factors calculated in different works, which may choose different normalisation.

    We then sum over all electrons to get the total atomic excitation factor, $K$, which we can use to reach the velocity-averaged differential cross section, expressed as
    \begin{equation}
        \frac{ \langle \mathrm{d} \sigma v \rangle }{ \mathrm{d} E } = 
         \frac{ \bar{\sigma}_e\,c }{ 2 m_e c^2 }
         \int \mathrm{d} v \frac{ f{(v)} }{ v/c }
         \int_{q_-}^{q_+} \, a_0^2 \, q \mathrm{d} q \,
         |F_{\chi}^{\mu}{(q)}|^2 K{(E,q)} \, ,
    \label{eqn:dsvdE}
    \end{equation}
    where $\bar{\sigma}_e$ is the free electron cross section at a fixed momentum transfer of $q=a_0^{-1}\approx 3.6\,{\rm keV}$, which we introduce following Ref.~\cite{Essig2012FirstXENON10,Essig2017} to simplify comparisons between results ($a_0$ is the Bohr radius; see Ref.~\cite{Roberts2019Electron-interactingDetectors} for full expression linking $\bar{\sigma}_e$ back to $\alpha_\chi$ parameter of Eq.~\eqref{eqn:yukawaV}). The DM speed distribution is denoted as $f$, which we assume to be a Maxwell-Boltzmann distribution with the Standard Halo Model assumptions (see, e.g.,Refs.~\cite{Nesti2013The2013,Freese2013iColloquium/iMatter}). 
    We note that electron recoil spectrum may be particularly sensitive to details of the velocity disctributions, see, e.g., Refs.~\cite{Roberts2019Electron-interactingDetectors,Radick2021DependenceDistribution,Roberts2016DarkSignal}.
    Lastly, $F_{\chi}$ is the DM form factor.

    As $F_{\chi}$ is able to be separated from the atomic excitation factor, the atomic excitation factors are largely independent of the DM model, and we only need to consider the electron coupling. For the example calculations in Sec.\ \ref{sec:egeventrate}, we use a DM form factor that is applicable to both vector and scalar couplings, as defined in Ref.\ \cite{Roberts2019Electron-interactingDetectors} as
    \begin{equation}
        F_{\chi}{(q)}
        =
        \frac{ (m_v / m_e)^2 + \alpha^2 }{ (m_v / m_e)^2 + (\alpha a_0 q)^2 } \, .
    \label{eqn:Fchi}
    \end{equation}

    While the effective potential that we began with can be applied to vector and scalar interactions, Eq.~(\ref{eqn:Knk_mel}) is only applicable to the vector case. We can reach scalar, pseudoscalar, or pseudovector electron couplings if we replace $e^{i \boldsymbol{q} \cdot \boldsymbol{r}}$ in the matrix element with $\gamma^0 e^{i \boldsymbol{q} \cdot \boldsymbol{r}}$, $\gamma^0 \gamma_5 e^{i \boldsymbol{q} \cdot \boldsymbol{r}}$, or $\gamma_5 e^{i \boldsymbol{q} \cdot \boldsymbol{r}}$, respectively.

%======================================================
\section{Calculations} \label{sec:calculations}
%======================================================

    %======================================================
    \subsection{Hartree-Fock approximation} \label{ssec:hf}

    As finding the atomic excitation factor includes a matrix element calculation with both the bound and continuum wavefunctions, we need to accurately form the electron orbitals.
    We use the relativistic Hartree-Fock (HF) method, which is both self-consistent and includes the electron exchange interaction. The final HF potential is then used to solve the Dirac equation at each step in a range of energy deposition values for the continuum wavefunctions. We write the HF potential as the sum of the direct and exchange potentials,
    \begin{multline}
        \hat{V}_{\mathrm{HF}} \psi_a \left( \v{r}_1 \right) = \sum_{i \neq a}^{N_c}
            \Bigg( \Bigg. 
                \int \frac{ \abs{ \psi_i (\v{r}_2) }^2 }{ r_{12} } \mathrm{d}^3 \v{r}_2 \, \psi_a(\v{r}_1)
                \\
                - \int \frac{\psi_i^{\dagger} (\v{r}_2) \psi_a (\v{r}_2)}{ r_{12} } \mathrm{d}^3 \v{r}_2 \, \psi_i (\v{r}_1)
            \Bigg. \Bigg) ,
    \label{eqn:V_HF}
    \end{multline}
    where the first term corresponds to the direct potential, the second term to the exchange potential, $i$ denotes the bound electron state with quantum numbers $\{ n_i, \kappa_i, m_i \}$, $N_c$ is the total number of electrons, and finally $r_{12} = \abs{\v{r}_1 - \v{r}_2}$.

    First, the Hartree-Fock equations are solved self-consistently for the $N_c$ bound electrons. 
    Then, the wavefunctions for the continuum electrons are found in the frozen Hartree-Fock potential by directly solving the Dirac equation, including the exchange term.
    (A small deviation from the frozen potential, known as the hole-particle interaction, will be discussed below.)
    We use the energy normalisation condition for the continuum states~\eqref{eq:e-norm}.
    In practice, this is achieved by solving the Dirac equation by integrating outwards to very large distances from the nucleus, where the wavefunctions take a spherical wave form analogous to the hydrogen-like case, and comparing to analytic solutions~\cite{Bethe1957QuantumAtoms}.

    %======================================================
    \subsection{Calculation of atomic ionization factors} \label{ssec:Kion}

    In the Dirac basis, we write the bound state orbitals as
    \begin{equation}
        %\braket{\boldsymbol{r}}{n\kappa m} 
        \phi_{n\kappa m}(\boldsymbol{r}) = \frac{1}{r}
        \left( \begin{matrix}
                f_{n\kappa}{(r)} \Omega_{\kappa m}{(\boldsymbol{\hat{n}})} \\
                ig_{n\kappa}{(r)} \tilde{\Omega}_{\kappa m}{(\boldsymbol{\hat{n}})}
        \end{matrix} \right) \, ,
    \label{eqn:boundstate_diracbasis}
    \end{equation}
    and similarly for the the continuum state orbitals
    \begin{equation}
        %\braket{\boldsymbol{r}}{\varepsilon\kappa m} 
        \phi_{\varepsilon\kappa m}(\v{r}) = \frac{1}{r}
        \left( \begin{matrix}
                f_{\varepsilon\kappa}{(r)} \Omega_{\kappa m}{(\boldsymbol{\hat{n}})} \\
                ig_{\varepsilon\kappa}{(r)} \tilde{\Omega}_{\kappa m}{(\boldsymbol{\hat{n}})}
        \end{matrix} \right) \, ,
    \label{eqn:contstate_diracbasis}
    \end{equation}
    where $f$ and $g$ are the large and small components of the  Dirac wavefunction, and $\Omega$ is a two-component spherical spinor. We can express $\Omega$ as
    \begin{equation}
        \Omega_{\kappa m} = \sum_{s_z} \braket{l, m-s_z, 1/2, s_z}{j, m}
         Y_{l, m-s_z}{(\boldsymbol{\hat{n}})} \chi_{s_z} \, ,
    \label{eqn:Omega}
    \end{equation}
    where $\braket{j_1 m_1 j_2 m_2}{J M}$ denotes a Clebsch-Gordon coefficient, $Y$ is a spherical harmonic, $\chi$ is a spin eigenstate, and $s_z$ is the electron spin, such that the sum runs over $s_z = \{-1/2,1/2\}$. Lastly, $\tilde{\Omega}$ is related to $\Omega$ through
    \begin{equation}
        \tilde{\Omega}_{\kappa m} 
        =
        - \left( \boldsymbol{\sigma} \cdot \boldsymbol{\hat{n}} \right) \Omega_{\kappa m}
        =
        \Omega_{-\kappa, m} \, .
    \label{eqn:Omegatilde}
    \end{equation}

    To calculate the matrix element in Eq.~(\ref{eqn:Knk_mel}), we use irreducible spherical tensors to expand the exponential term:
    \begin{equation}
        e^{i \boldsymbol{q} \cdot \boldsymbol{r}} = \sum_{L=0}^{\infty} \sum_{M=-L}^{L} T_{LM} \, ,
    \label{eqn:eiqrexpansion}
    \end{equation}
    where
    \begin{equation}
        T_{LM} = 4\pi i^L j_L{(qr)} Y_{LM}{(\theta_r, \phi_r)} Y^*_{LM}{(\theta_q,\phi_q)} \, .
    \label{eqn:TLMtensor}
    \end{equation}
    where $j_L$ is a spherical bessel function of the first kind and $L$ is the multipolarity. With this, and angular reduction rules \cite{Varshalovich1988QuantumMomentum}, we can then write the atomic excitation factor as
    \begin{equation}
        K_{n\kappa }{(E,q)} = E_H \sum_L \sum_{\kappa'}
         | R_{n\kappa}^{\kappa' L} |^2 C_{\kappa \kappa'}^{L} \, ,
    \label{eqn:Knk_RC}
    \end{equation}
    where $R$ is the radial integral and $C$ is an angular coefficient. In the case of a vector electron coupling, $R$ can be expressed as
    \begin{equation}
        R_{n\kappa }^{\kappa' L} = \int_0^{\infty} \left[ f_{n\kappa}{(r)} f_{\varepsilon \kappa'}{(r)}
         + g_{n\kappa}{(r)} g_{\varepsilon \kappa'}{(r)} \right] j_L{(qr)} dr \, .
    \label{eqn:RnkkL_vector}
    \end{equation}

    If we instead consider the scalar case, where we replace $e^{i \boldsymbol{q} \cdot \boldsymbol{r}}$ with $\gamma^0 e^{i \boldsymbol{q} \cdot \boldsymbol{r}}$, this will result in a radial integral of
    \begin{equation}
        R_{n\kappa }^{\kappa' L} = \int_0^{\infty} \left[ f_{n\kappa}{(r)} f_{\varepsilon \kappa'}{(r)}
         - g_{n\kappa}{(r)} g_{\varepsilon \kappa'}{(r)} \right] j_L{(qr)} dr \, ,
    \end{equation}
    which can be re-written in a form that is more numerically stable at low $q$ due to the orthogonality condition:
    \begin{equation}
    \begin{split}
        R_{n\kappa }^{\kappa' L} = \int_0^{\infty} \Big[ f_{n\kappa}{(r)} f_{\varepsilon \kappa'}{(r)}(j_L{(qr)}-1)
         \\- g_{n\kappa}{(r)} g_{\varepsilon \kappa'}{(r)}(j_L{(qr)}+1) \Big]  dr \, ,
         \end{split}
    \end{equation}
    as discussed below.
    For the pseudoscalar case, replacing $e^{i \boldsymbol{q} \cdot \boldsymbol{r}}$ with $\gamma^0 \gamma_5 e^{i \boldsymbol{q} \cdot \boldsymbol{r}}$ gives
    \begin{equation}
        R_{n\kappa }^{\kappa' L} = \int_0^{\infty} \left[ f_{n\kappa}{(r)} g_{\varepsilon \kappa'}{(r)}
         + g_{n\kappa}{(r)} f_{\varepsilon \kappa'}{(r)} \right] j_L{(qr)} dr \, ,
    \end{equation}
    and finally, for the pseudovector case, replacing $e^{i \boldsymbol{q} \cdot \boldsymbol{r}}$ with $\gamma_5 e^{i \boldsymbol{q} \cdot \boldsymbol{r}}$ gives
    \begin{equation}
        R_{n\kappa }^{\kappa' L} = \int_0^{\infty} \left[ f_{n\kappa}{(r)} g_{\varepsilon \kappa'}{(r)}
         - g_{n\kappa}{(r)} f_{\varepsilon \kappa'}{(r)} \right] j_L{(qr)} dr \, .
    \end{equation}

    In the cases of both vector and scalar electron couplings, $C$ can be expressed as
    \begin{equation}
        C_{\kappa\kappa'}^{L} = 
         [j][j'][L]
         \left( \begin{matrix}
                 j & j' & L \\
                 -1/2 & 1/2 & 0
         \end{matrix} \right)^2
         \Pi_{ll'}^{L} \, ,
    \end{equation}
    where $[J] \equiv 2J+1$, the term in parenthesis is a Wigner 3-j symbol, and $\Pi_{ll'}^L$ the parity selection rule (it is unity if $l+l'+L$ is even and zero otherwise).
    The angular coefficient for the pseudovector and pseudoscalar cases are similar, but we replace $\kappa$ with $\tilde{\kappa} = -\kappa$ and $l$ with $\tilde{l} = |\tilde{\kappa} + 1/2| - 1/2$.

\begin{figure}
        \centering
        \includegraphics{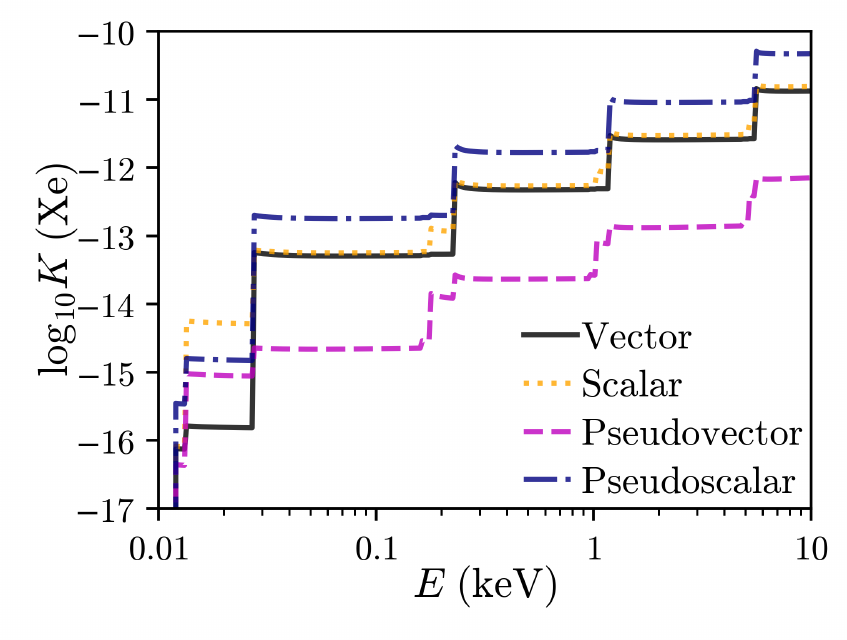}
        \caption{Comparison of the total atomic excitation factor as a function of energy deposition for xenon for vector (solid line), scalar (dotted line), pseudovector (dashed line), and pseudoscalar (dash-dotted line) electron couplings at a fixed momentum transfer of $q=4$ MeV.}
    \label{fig:Xe_coupl_Ktot_vs_E}
    \end{figure}

\begin{figure}
        \centering
        \includegraphics{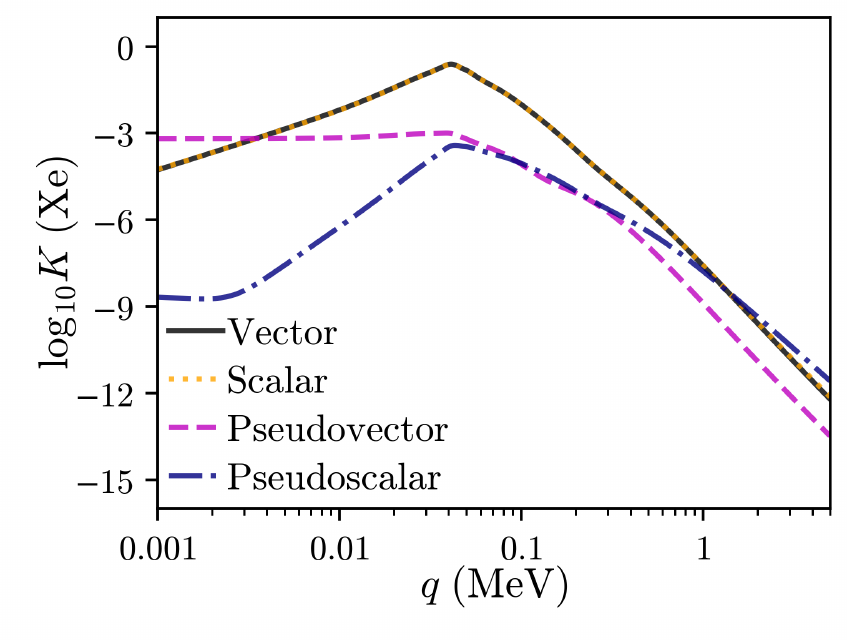}
        \caption{Total atomic excitation factors as a function of momentum transfer for xenon at a fixed energy deposition of $E=2$ keV, with the same electron couplings as in Fig.~\ref{fig:Xe_coupl_Ktot_vs_E}.}
    \label{fig:Xe_coupl_Ktot_vs_q}
    \end{figure}

    Example calculations for the total atomic factors for each of these couplings can be seen in Figs.~(\ref{fig:Xe_coupl_Ktot_vs_E}) and (\ref{fig:Xe_coupl_Ktot_vs_q}) as functions of energy deposition and momentum transfer, respectively. Tables of these factors are included as supplementary material. Alternatively, the tables can be found on GitHub \cite{Roberts2023AtomicIonisation}, alongside example code that uses the tables to calculate cross sections as seen in Sec.~\ref{sec:electronimpact} and Sec.~\ref{sec:egeventrate}.

    As a stability test for the calculations, we also computed $K$ using an approximate local potential instead of the HF potential. The local potential used was the nuclear potential plus a parametric potential, making the calculation far simpler and less prone to numerical instabilities. 
    The local potential leads to very similar atomic excitation factors, giving a point of comparison that highlights possible numerical issues or errors in the calculation.

        \begin{figure}%[h!]
            % \centering
            % \raggedleft
            \includegraphics[left]{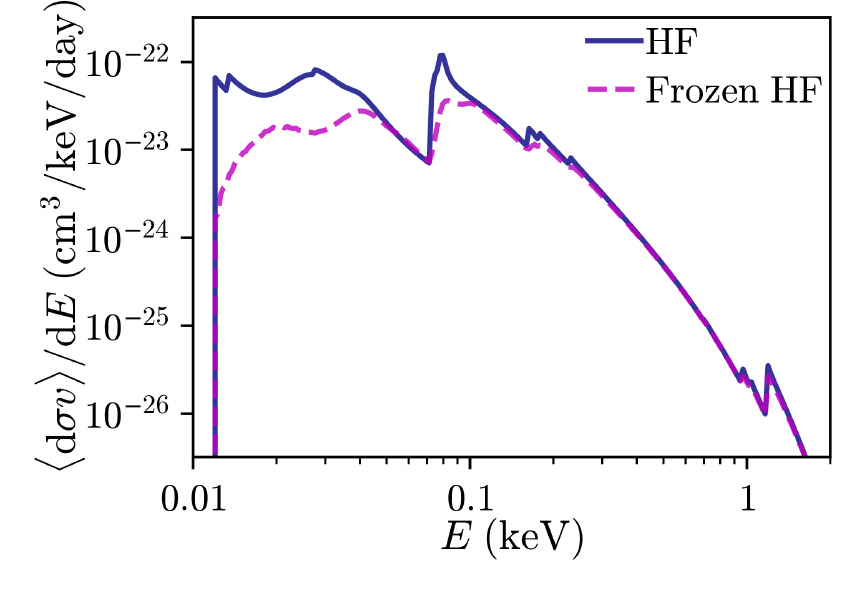}
            \caption{Velocity-averaged differential cross section for xenon when  accounting for the hole-particle interaction (solid line),
            and when excluding it (``frozen'' Hartre-Fock, dashed line). In this case, we have set the DM mass to be $m_{\chi} = 1$ GeV. The DM form factor is set to $F_{\chi}=1$, corresponding to a heavy mediator as per Eq.~\ref{eqn:Fchi}. We use a vector electron coupling with free electron cross-section $\bar{\sigma}_e = 10^{-37}$ cm$^2$.}
            \label{fig:Xe_dsvde_hpcomp}
        \end{figure}

    %======================================================
    \subsection{Many-body effects} \label{ssec:many-body}

                \begin{figure}
            \includegraphics[width=0.3\linewidth]{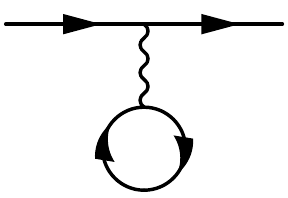}~
            \includegraphics[width=0.3\linewidth]{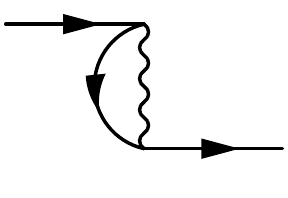}~
            \includegraphics[width=0.3\linewidth]{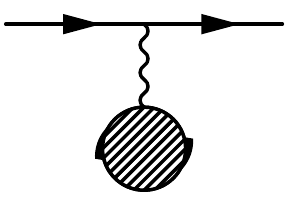}
            \caption{Goldstone diagrams for the direct (left), exchange (middle), and hole-particle (right) contributions to the atomic potential. 
            The wavy line is the Coulomb interaction, external lines are either bound atomic states or unbound continuum states, internal lines are core states (holes).
            The hole-particle effect arises to the deviation of the direct potential for the ejected continuum electron from that of the bound electrons.}
            \label{fig:hp}
        \end{figure}

        The most important many-body effect (that is, deviation from the frozen-core Hartree Fock approximation) is the hole-particle interaction.
        Physically, this effect arises due to the 
        deviation of the (direct) Hartree-Fock potential for the ionized electron compared to those in the core, as in Fig.~\ref{fig:hp}.
        In practical Hartree-Fock calculations for occupied core states, the self-interaction term is included in the direct potential; this is then exactly compensated by the corresponding term in the exchange potential, e.g., by  setting $i=a$ in Eq.~(\ref{eqn:V_HF}). 
        However, this cancellation does not apply to an electron that has been excited out of the core.
        Therefore, the self-interaction term should be removed for the excited states.
        While this hole-particle interaction term makes a very small difference when looking at the ionization of bound electrons with high ionization energies (close to the nucleus), the impact becomes more obvious for electrons further from the nucleus, and for lower-energy scatterings. 
        From Fig.~\ref{fig:Xe_dsvde_hpcomp}, we can see that this contribution to the atomic factors carries through to the cross section calculation, resulting in a more significant discrepancy as we move to lower energies.

        \begin{figure}
            % \centering
            % \raggedleft
            \includegraphics[width=0.4\linewidth]{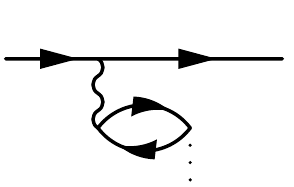}~
            \includegraphics[width=0.4\linewidth]{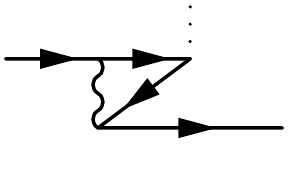}\\
            \includegraphics[width=0.4\linewidth]{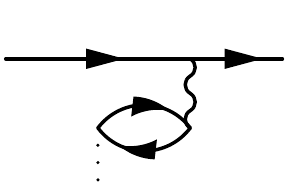}~
            \includegraphics[width=0.4\linewidth]{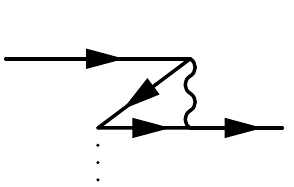}
            \caption{Goldstone diagrams for core polarisation correction to matrix elements, showing the direct (left) and exchange (right) contributions.
            The diagram notation is the same as in Fig.~\ref{fig:hp}, with the dotted line representing interaction with the external field (DM particle), and the internal forward lines representing virtual excited atomic states.
           In the all-order RPA method, each of the external field vertices are replaced with these four diagrams; this process is repeated until the matrix elements converge.}
            \label{fig:rpa}
        \end{figure}

        Beyond Hartree-Fock and hole-particle effects, the next most important many-body correction to matrix elements of the external field is the {\em core polarisation}, often referred to as the relativistic random phase approximation with exchange (RPA)~\cite{Johnson1980RelativisticApproximation}.
        The lowest-order core polarisation diagrams are shown in Fig.~\ref{fig:rpa}.
        Physically, this effect arises due to the combined action of the external field and inter-electron Coulomb interaction.
        In the present case, it manifests in the possibility that when the dark matter particle interacts with an electron in one state, an electron in a different state may become ionized due to its Coulomb interaction with the first electron.
        This may be particularly important in cases where, for example, only $s$-states have appreciable $e^{iqr}$ matrix elements (due to their non-zero wavefunctions at the nucleus), but they are energetically inaccessible. In that case, an energetically accessible $p,d$-state may be ejected via Coulomb interaction with an $s$-electron that interacts with the dark matter.
        
        In the many-body perturbation theory diagrams, there is an implied sum over intermediate states, with the backward lines denoting the core (bound, occupied) atomic states, and the forward lines representing the full spectrum of excited (unoccupied) bound and continuum states. To approximate this spectrum, we form an approximately complete, though finite, basis by diagonalising a set of B-splines over the atomic Hamiltonian (see, e.g., ~\cite{Johnson1988FiniteSplines,Shabaev2004DualEquation}). We use the dual-kinetic-balance basis as introduced in Ref.~\cite{Beloy2008ApplicationStructure}, which offers a high level of convergence and stability. 
        Our technique is well tested, and produces high-accuracy results across a range of atomic systems~\cite{Roberts2022ElectricElectron}.
        
        The lowest-order calculation typically overestimates the core-polarisation contribution, and often significantly.
        For accurate calculations, all-orders (in the Coulomb interaction) calculations are required.
        In the RPA approximation, this is achieved iteratively, by replacing each external field vertex in the four RPA diagrams (Fig.~\ref{fig:rpa}) with the four RPA diagrams. This process is continued iteratively until convergence is reached.
        This is equivalent (up to first-order in external field) to including the action of the external field into the potential when solving the Hartree-Fock equations (known as time-dependent Hartree-Fock method~\cite{Dzuba1984RelativisticAtoms}).
        We note that the RPA equations must be iterated separately for each value of momentum transfer $q$, and each multipolarity, $L$ (see Eq.~\eqref{eqn:eiqrexpansion}). We thus include 
        %only lowest-order core=polarisation effects, and calculated 
        the all-orders RPA effects only for a subset of the parameter range to check its contribution, which we find to be small.

        The lowest-order core polarisation effects has the largest impact at small energy deposition values, and gives at most a correction of a factor of two.
        The all-orders RPA corrections significantly reduce this. After integrations, the uncertainty in Hartree-Fock calculations from excluding RPA is of order $\simeq20\%$, which is more than sufficient for the current purpose.

        As a final consideration, we check for errors due to imperfect orthogonality.
        Due to numerical uncertainties, the exact orthogonality is not guaranteed between
        the core and continuum states in practical calculations, particularly when the hole-particle interaction is included.
        This can be source of significant errors, so must be checked.
        We can check for this by explicitly enforcing orthogonality between the continuum states and
        the states in the core using the standard procedure:
        $\ket{a} \rightarrow \ket{a} - \ket{b}\braket{b}{a}$.
        In our calculations, this makes very little difference, due to the already good orthogonality achieved in
        the self-consistent Hartree-Fock procedure.
        However, in methods where the self-consistency of the potential cannot be guaranteed, this check is essential.

        Another option for ensuring orthogonality between core and continuum states is to make a substitution in the matrix element of $e^{i \boldsymbol{q} \cdot \boldsymbol{r}} \rightarrow e^{i \boldsymbol{q} \cdot \boldsymbol{r}} - 1$ in Eq.~\ref{eqn:Knk_mel} (or, equivalently, $j_L \rightarrow j_L - 1$ in Eq.~\ref{eqn:Knk_RC}). While this corrects errors caused by orthogonality in regions of low momentum transfer, it can shift the error to high momentum transfer. As we need accurate atomic factors across many orders of magnitude of momentum transfer for DM-electron scattering, this substitution alone cannot ensure orthogonality for this case. This is especially true when accounting for the hole-particle interaction. As a result, we opt to use the previous procedure for ensuring orthogonality, as this addresses issues across the required range.

    %======================================================
    \subsection{Approximation of Atomic Excitation Factors} \label{ssec:approximationofK}

        We present an informative approximate approach to presenting the ionization form factors that is valid for $q\gg 1/a_0$.
        When the energy deposition goes above the ionization of a bound electron, this electron becomes `accessible'; the atomic excitation factor for that electron is also zero below this point. For an increasing energy deposition past this point, the atomic excitation factor is relatively constant while the continuum energy remains small, so long as $q$ is reasonably large. 
        This is because for large $q$, only the low-$r$ part of the electron wavefunction may contribute. In this region, the energy of the bound or continuum electron is insignificant compared to the nuclear potential $|\en|\ll|Z/r|$, and the Dirac equation is independent of energy.
        In this case, we may approximate $K_{n\kappa}{(E,q)}$ as a step function of energy, allowing it to be expressed as
        \begin{equation}
            K_{n\kappa}{\left( E,q \right)}
            \approx
             \tilde{K}_{n\kappa}{\left( q \right)} 
             \Theta{\left( E - I_{n\kappa} \right)} \, ,
        \label{eqn:Knk_step}
        \end{equation}
        where $\Theta$ is the Heaviside step function, and $\tilde{K}_{n\kappa}$ is the atomic excitation factor that is dependent on momentum transfer at a fixed energy deposition. Thus, above the ionization energy, $K_{n\kappa}$ is equal to $\tilde{K}_{n\kappa}.$

        This approximate method loses accuracy for very small values of momentum transfer ($q \lesssim 0.1$ MeV). For the typical momentum transfer values for DM-electron scattering, inaccuracies found in the low-$q$ region do not have a significant impact on cross-section calculations due to integrating over $q$, as seen in Eq.~(\ref{eqn:dsvdE}), but care should be taken if high accuracy is needed in this region.

%======================================================
\section{Electron Impact ionization} \label{sec:electronimpact}
%======================================================

    We have also calculated the total cross section for the case of an atomic electron being ionized by an incoming free electron (referred to as `electron impact' (EI) ionization in the literature). With existing experimental data, the calculations being applied to this interaction type serves as a test of the accuracy of the code. 
    In the case that $m_\chi\to m_e$, $m_v\to0$, and $\alpha_\chi\to\alpha$, it can be seen that the electron-impact and DM-induced ionizations are very similar atomic processes.

    To compare to experimental data, we can calculate the total cross section,
    \begin{equation}
        \sigma{(E_i)} = \frac{4\pi}{E_i} \int_{0}^{E_i} \int_{q_-}^{q_+} \mathrm{d}q \frac{K{(q,E)}}{q^3} \mathrm{d}E \, ,
    \label{eqn:totxsection_impact}
    \end{equation}
    where $E_i$ is the incident energy of the projectile electron, and $E$ is again the energy deposition.

    \begin{figure}
        \centering
        \includegraphics[left]{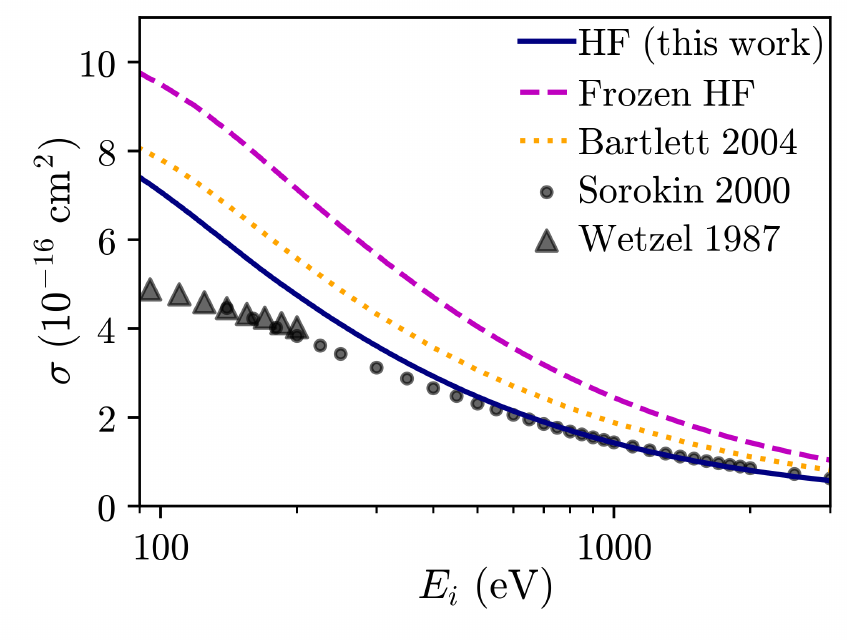}
        \caption{Calculations of total electron impact ionization cross sections  for xenon with (HF) and without (Frozen HF) accounting for the hole-particle interaction, and comparison to calculations by Bartlett {\em et al.}~\cite{Bartlett2004Electron-impactZ=54} and experimental results from Sorokin {\em et al.}~\cite{Sorokin2000MeasurementsPhotoionization} and Wetzel {\em et al.}~\cite{Wetzel1987AbsoluteMethod}.}
    \label{fig:Xe_stot_vs_E_highE}
    \end{figure}

    As found in past literature, calculations of the total cross sections for this type of scattering are, on average, overestimated when compared to experimental values (see, e.g., Ref.~\cite{Bartlett2002CalculationSections} and references therein). 
    For testing DM-induced ionization, the relevant incident energy scale is $m_\chi v^2\sim({\rm GeV})(10^{-3})^2\sim {\rm keV}$.
    Our calculations are presented on Fig.~\ref{fig:Xe_stot_vs_E_highE}, along with experimental data, and existing calculations for comparison.
    
    When the incident electron energy is low ($E_i \lesssim 100$ eV), we can see that our calculations (and those of other groups) overesimate the total ionization cross section.  This divergence from experiment largely stems from our use of the Born approximation \cite{Bartlett2004Electron-impactZ=54}, which is only valid for very weak interactions. As the DM-electron interactions that we consider in this work falls into this category \cite{Roberts2016IonizationMatter}, the approximation holds, and does not carry the same importance as it does when looking at electron impact ionization.
    Further, the electron-impact ionization case is complicated by the presence of an exchange term in the interaction (which we did not account for), since the projectile is an electron which may be exchanged with a bound atomic electron. Since no such term exists in the DM case, the accuracy for the DM scattering case is expected to be higher.

    Importantly, in the region where energies are closer to that of an DM-electron scattering event ($E_i \gtrsim 1$ keV), we can see in Fig.~\ref{fig:Xe_stot_vs_E_highE} that the case where we have subtracted the hole-particle interaction gives results that are closest to experiment. This shows an improvement on the accuracy of existing results in this high-energy region and highlights the importance of subtracting the hole-particle interaction term.
    This is both experimental verification of our atomic wavefunctions, and also of the approximations used to derive the scattering formula.

%======================================================
\section{Example Event Rate Calculation} \label{sec:egeventrate}
%======================================================
    
    For DM-electron scattering, the differential event rate is,
    \begin{equation}
        \frac{ \mathrm{d} R }{ \mathrm{d} E }
        =
        \frac{n_T \rho_{\chi}}{m_{\chi}c^2} 
        \frac{ \langle \mathrm{d} \sigma v \rangle }{ \mathrm{d} E } \, ,
    \label{eqn:dRdE_erate}
    \end{equation}
    where $n_T$ is the number of target atoms in the detector per unit mass (reciprocal of atomic mass), $m_{\chi}$ is the DM mass, and $\rho_{\chi}$ is the local energy density of DM, which we take to be $\sim 0.4 \, \mathrm{GeV \ cm}^{-3}$ \cite{Bovy2012ONDENSITY}. However, we need to consider that not all events are feasibly detectable and account for capability of the detector itself. This allows us to find observable event rates for specific energy regions and then compare to the event rates seen in experiments.

    As the detection capabilities vary for each detector depending on the equipment, the set-up of the detector and the scintillating material, the model will also change per experiment. As an example of this calculation, we follow Ref.~\cite{Aprile2020} and start by modelling the detector resolution of XENON1T as a Gaussian with standard deviation,
    \begin{equation}
        \sigma{(E)} = a \cdot \sqrt{E} + b \cdot E \, ,
    \label{eqn:sigma}
    \end{equation}
    where $a=\left(0.310 \pm 0.004\right) \sqrt{\mathrm{keV}}$ and $b=0.0037 \pm 0.0003$. We use this Gaussian, $g_{\sigma}$, to smear the theoretical event rate, given in Eq.~(\ref{eqn:dRdE_erate}).

    The event rate also needs to be corrected for the efficiency of the detector, which we accounted for by fitting the total efficiency as a function of energy, $\varepsilon{(E)}$, as given in Fig.~2 of Ref.~\cite{Aprile2020}. 
    Combining these steps together, we can express the observable differential event rate as
    \begin{equation}
        \frac{\mathrm{d}S}{\mathrm{d}E} = \varepsilon{(E)} \int_{0}^{\infty}
         g_{\sigma}{\left( E' - E \right)}
         \frac{ \mathrm{d} R{(E')}}{\mathrm{d} E'} \mathrm{d}E' \, .
    \label{eqn:dSdE_erate}
    \end{equation}

    \begin{figure}
        \centering
        \includegraphics{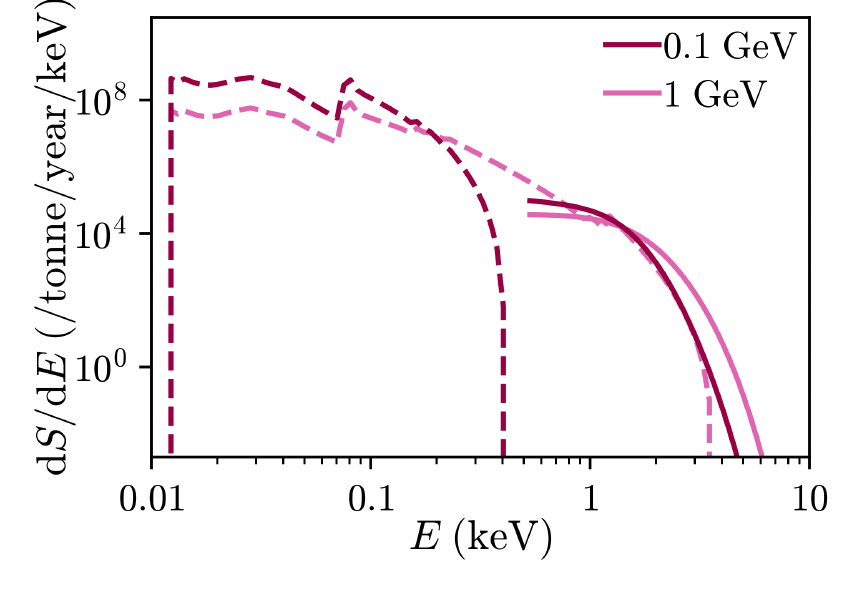}
        \caption{Example event rate calculations before (dotted lines) and after (solid lines) smearing with the Gaussian and correcting for detection efficiency~\cite{Aprile2020},
        corresponding to Eqs.~(\ref{eqn:dRdE_erate}) and (\ref{eqn:dSdE_erate}), respectively. 
        The spikes seen in the calculated event rate are due to deeper shells becoming accessible as the energy deposition increases, which we can also see in the cross section. A cut-off has been applied to the observable event rate at $0.5$ keV to indicate the minimum energy threshold. Again, $\bar{\sigma}_e = 10^{-37}$ cm$^2$.}
        \label{fig:dSdE_dRdE_comp}
    \end{figure}

    From Fig.~\ref{fig:dSdE_dRdE_comp}, we can see that correcting for the specific detector dampens the event rate as expected. For the lowest mass case, $0.1$ GeV, the calculated event rate drops to zero at a lower energy than the observable event rate peaks at. This is due to the detector response being modelled as a Gaussian, which allows very low energy signals to `leak' through to higher energy regions and be considered detectable.
    While this is plausible at higher energies \cite{Aprile2020EnergyRange}, this introduces large uncertainties in the observable event rate at low energies, where it's not appropriate to model the response using a simple Gaussian.

    From Fig.~\ref{fig:dSdE_dRdE_comp}, we can see that the theoretical event rate peaks below $1$ keV for all DM mass cases. Although the tail ends of each Gaussian will have very low magnitude in this region, the magnitude of the event rate causes the observable event rate to be high as well. 
    This implies that modelling the detector in this way overestimates the observable event rate when looking at any scattering type that has large contributions in regions of low energy deposition. In these cases, care should be taken to determine the strength of the effect that the use of the Gaussian has on the observable event rate, and whether an overestimation is being introduced by events happening far below threshold. If the effect is large, a more accurate option is to directly simulate the specific detector, such as the use of NEST (Noble Element Simulation Technique) \cite{Szydagis2011NEST:Xenon,Szydagis2013EnhancementXenon,Mock2014ModelingNEST,Lenardo2015AXenon}, or Obscura~\cite{Emken2021Obscura:Recoils}.
    This is particularly important to consider for scattering cases involving DM, where the nature of the particle is unknown, as an overestimation in the predicted event rate could lead to DM models being excluded preemptively.

    \begin{figure}
        \centering
        \includegraphics{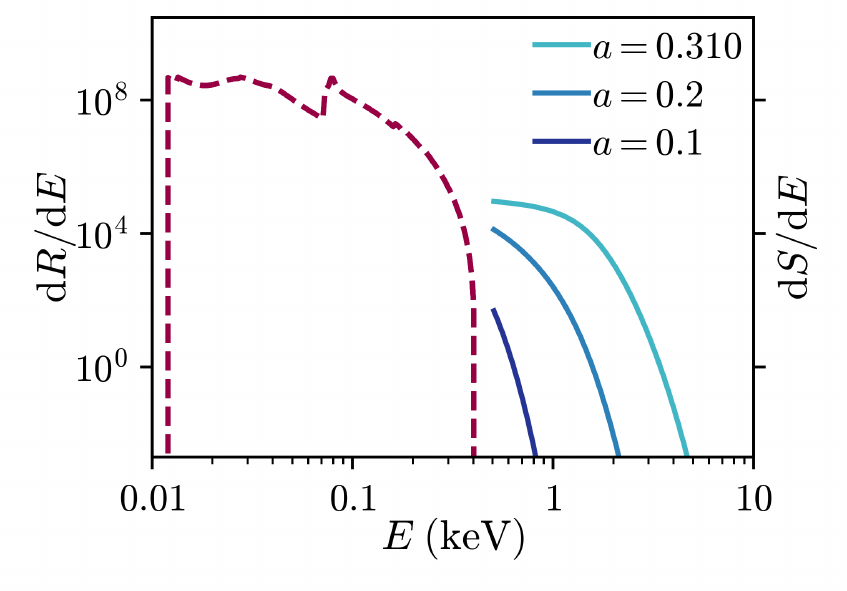}
        \caption{Example rate calculations for $m_{\chi}=0.1$ GeV, with a vector electron coupling and a heavy mediator. The scattering rate (dashed line) represents the case of a perfect detector, matching the theoretical $\mathrm{d}R/\mathrm{d}E$, whereas the observable event rate (solid lines) represent the rate that would be seen by a detector with varying energy resolution modelling. The case of $a=0.310$ matches that of XENON1T, while $a=0.2$ and $a=0.1$ are used as comparison to highlight the sensitivity to any change in the Gaussian. Again, $\bar{\sigma}_e = 10^{-37} \mathrm{cm}^2$.}
        \label{fig:my_label}
    \end{figure}
    
%======================================================
\section{Conclusion} \label{sec:conclusion}

    We have presented tables of atomic excitation factors as functions of momentum transfer and energy deposition for DM-electron interactions with vector, scalar, pseudovector, and pseudoscalar couplings for argon, krypton, and xenon \cite{Roberts2023AtomicIonisation,SeeFactors}. These take relativistic effects into account and provides an accurate depiction of the atomic physics involved. As such, they can be combined with a DM model of choice to calculate cross sections and event rates, without risking the underestimates that are common when neglecting the atomic physics.

    We have tested the code for any numerical instabilities and errors by calculating the atomic factor with an approximate potential. We have also tested the code for accuracy by applying the calculations to electron impact ionization and found good agreement with experimental results.

    As these atomic factors may be used to calculate cross sections and event rates, we have presented an example of an event rate calculation specific to the XENON1T experiment. While accurate atomic physics is necessary to reach reliable event rates, we have also highlighted the importance of the modelling of the detector, particularly the low-energy response.

% %======================================================
\acknowledgements
This work was supported by the Australian Research Council through DECRA Fellowship No.\ DE210101026, and Grants No.\ DP230101058 and DP200100150.
We thank Ben Carew for discussions.

\bibliography{references.bib}
% \bibliographstyle{apsrev4-2}

\end{document}